\documentstyle[psfig]{mn}

\input{epsf}

\begin{document}

\title{Cross correlations of X-ray and optically selected clusters with near
infrared and optical galaxies}
\author[A. G. S\'anchez et al.]
{\parbox[t]{\textwidth}{
Ariel G. S\'{a}nchez$^{1,2}$\thanks{E-mail: arielsan@oac.uncor.edu}, 
Diego G. Lambas$^{1,2}$, 
Hans B\"{o}hringer$^3$, 
Peter Schuecker$^3$.} 
\vspace*{6pt} \\
$^1$ Grupo de Investigaciones en Astronom\'{\i}a Te\'{o}rica y Experimental,
 (IATE), Observatorio Astron\'{o}mico C\'{o}rdoba, UNC, Argentina.\\
$^2$ Consejo Nacional de Investigaciones Cient\'{\i}ficas y Tecnol\'ogicas 
(CONICET), Argentina.\\
$^3$ Max-Planck-Intitut fur Extraterrestriche Physik, P.O. Box 1312, 85741 
Garching, Germany.\\
}

\date{8 March 2005}
\maketitle

\begin{abstract}

We compute the real-space cluster-galaxy cross-correlation 
$\xi _{cg}(r)$ using the ROSAT-ESO Flux Limited X-ray (REFLEX) cluster 
survey, a group catalogue constructed from the final version of the 2dFGRS, 
and galaxies extracted from 2MASS and APM surveys. This first 
detailed calculation of the cross-correlation for X-ray clusters and 
groups, is consistent with previous works and shows that $\xi _{cg}(r)$ can not 
be described by a single power law. We analyse the clustering dependence 
on the cluster X-ray luminosity $L_{X}$ and virial mass $M_{vir}$ thresholds 
as well as on the galaxy limiting magnitude. We also make a comparison 
of our results with those obtained for the halo-mass cross-correlation 
function in a $\Lambda $CDM N-body simulation to infer the scale 
dependence of galaxy bias around clusters. Our results indicate that the 
distribution of galaxies shows a significant anti-bias at highly non-linear 
small cluster-centric distances ($b_{cg}(r)\simeq0.7$), 
irrespective of the group/cluster virial mass or X-ray luminosity and galaxy 
characteristics, which show that a generic process controls the efficiency 
of galaxy formation and evolution in high density regions. 
On larger scales $b_{cg}(r)$ rises to a nearly constant value of order 
unity, the transition occuring at approximately $2\;h^{-1}$Mpc for 2dF 
groups and $5\;h^{-1}$Mpc for REFLEX clusters. 
\end{abstract}


\section{Introduction}

\begin{figure*}
\centering
\centerline{\epsfysize = 12cm \epsfbox[20 160 580 670]{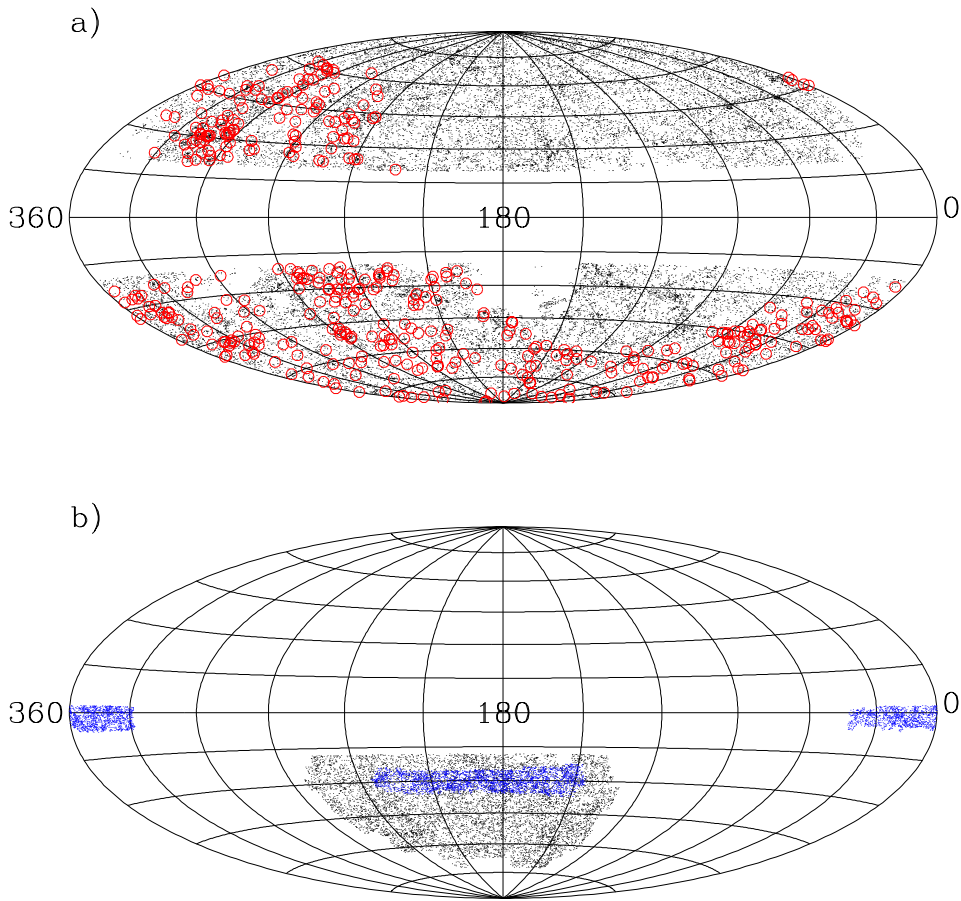}}
\caption{
a) Aitoff projection of the distribution, in galactic coordinates, of 2MASS 
galaxies with $K_s < 12.5$ (dots) together with the REFLEX cluster sample 
(red circles). The angular mask described in section 2.2 has been applied to 
both samples. b) Aitoff projection of the distribution, in equatorial 
coordinates, of 10\% of APM galaxies with $b_j < 18.5$ (black dots) 
and all 2dFGGC (blue dots).
}
\label{fig1}
\end{figure*}

\noindent Statistical analysis of clusters and groups of galaxies are among
the most powerful tools for the study of large scale structure. In the
hierarchical clustering scenario, these objects form by the gravitational
amplification of primordial density fluctuations, so that they can be used to
characterize the spatial distribution of the peaks of the density field.
Besides, as they are the largest gravitationally bound objects and span a
wide range of masses, they can be used as laboratories to determine the role
of the different astrophysical processes that govern galaxy formation.

The first statistical analysis on this type of objects where based on
cluster catalogues constructed by visual identification (Abell, 1958; Abell,
Olowin \& Corwin, 1989). The cluster-cluster correlation function $\xi
_{cc}(r)$ has been the preferred tool for such analysis, providing a simple
and convenient statistical tool to characterize their spatial distribution
and place constraints on cosmological models (Bahcall \& Soneira 1983;
Klypin \& Kopylov 1983; Peacock \& West 1992; Postman, Huchra \& Geller
1992). Nonetheless, clear evidence of inhomogeneities and line-of-sight
projection effects was found (Lucey 1983; Sutherland 1988), which where
responsible for part of the observed amplitude of the angular correlation.
The situation improved with digitized cluster surveys like the
Edinburgh/Durham Cluster Catalogue (Nichol et al. 1992) and the APM survey
(Dalton et al 1992, 1994; Croft et al 1997). Although these where more 
homogeneous, systematics due to projection effects are inherent to cluster 
identification based on angular positions in the sky (Padilla \& Lambas, 
2003).

In recent years, X-ray information has been used to construct new cluster
samples. This method of identification has several advantages with respect
to the use of optical data. The X-ray emission is a strong signature of a
gravitational potential well since there is a strong relation between the
X-ray luminosity $L_{X}$ and mass (Reiprich \& B\"{o}hringer 2002). Besides
the emissivity of thermal Bremsstrahlung radiation is proportional to the
square of the electron density thus it provides clear evidence of a mass
concentration which is not strongly affected by projection effects.
Furthermore, the X-ray emission from the clusters is concentrated towards
the dense central core giving an improved angular resolution compared to the 
one obtained using galaxy concentrations, also helping to reduce the possibility
of projection effects. The ROSAT-ESO Flux Limited X-ray (REFLEX) cluster
survey (B\"{o}hringer et al. 2004), comprising 447 objects, is the largest
statistically complete X-ray selected sample. It has been previously used to
analyse the large scale structure through its power spectrum (Schuecker et 
al. 2001) and its correlation function (Collins et al., 2000). Although 
Schuecker et al. (2001) found a systematic increase of the amplitude of 
the power spectrum with limiting X-ray luminosity, Collins et al. (2001) 
results  lack such a trend in the correlation function, characterized by 
an approximate power-law with correlation length $r_{0}=18.8 \pm 
0.9\;h^{-1}\rm{Mpc}$.

On the other hand, since the advent of large galaxy redshift surveys a
reliable identification of groups of galaxies has been possible using not 
only angular positions, but also the complete redshift information, 
avoiding in this way a strong contamination by projection effects.
The first algorithms designed for this task where developed by Huchra \&
Geller (1982) and Nolthenius \& White (1987). These methods where based on
the friends-of-friends algorithm used in numerical simulations with a 
redshift dependent linking length parameter. A variety of 
group catalogues where constructed using this algorithms: Merch\'{a}n et 
al. (2000) used the Updated Zwicky Catalogue (Falco et al. 1999), 
Giuricin et al. (2000) constructed a group catalogue from the Nearby 
Optical Galaxy Sample, Tucker et al. (2000) extracted a group sample 
from the Las Campanas Redshift Survey (Shectman et al. 1996), and 
Ramella et al. (2002) used the information of the Updated Zwicky
catalogue and the Southern Sky Redshift Survey (da Costa et al. 1998).
Recently, the 100K public data release of the 2dF Galaxy Redshift Survey 
(2dFGRS, Colles et al. 2001) has been used by Merch\'an \& Zandivarez (2002) 
to construct one of the largest group catalogues using a slight 
modification of the Huchra \& Geller (1982)
algorithm. The large scale redshift space distribution of this sample 
has been studied by Zandivarez, Merch\'{a}n \&
Padilla (2003). They showed that the redshift space correlation function has
a power law behavior with a correlation length of 
$s_{0}= 8.9\pm0.3\; h^{-1}\rm{Mpc}$. 

The cluster and group catalogues identified by X-ray information or 
three dimensional percolation algorithms, represent different types of 
objects. In order to be able to confront the results obtained for these 
samples with the predictions of theoretical models, a complete 
understanding of the differences in the statistical properties of the 
samples is crucial. One of the most important possible comparisons is the 
analysis of the clustering of galaxies, selected in different wave-bands, 
around these objects and how it depends on the properties like X-ray 
luminosity or virial mass. The best
statistical tool for such analysis is the two-point cluster-galaxy
cross-correlation function $\xi _{cg}(r)$, which is a measure of the mean
radial density profile around clusters. The first measurement of this
correlation function was performed by Seldner \& Peebles (1977), who studied
the angular cross-correlation of Abell clusters and Lick galaxies. They
found that $\xi _{cg}(r)$ was very large and positive out to
scales as large as $r\simeq 100\; h^{-1}\rm{Mpc}$ but they where 
contradicted by Lilje \& Efstathiou (1988) who did not find any strong 
evidence of clustering beyond $r>=20\; h^{-1}\rm{Mpc}$ arguing that part 
of the signal detected by Seldner \& Peebles (1977) was a consequence of 
artificial surface density gradients in the Lick catalogue. At smaller 
scales, Lilje \& Efstathiou (1988) found that $\xi _{cg}(r)$ was well 
fitted by a power law with $r_{0}=8.8\;h^{-1}\rm{Mpc}$.
Later works (Dalton 1992; Mo, Peacock \& Xia 1993; Moore et al. 1994) 
found similar results based on direct measurements from redshift
surveys of galaxies and clusters. They found that $\xi _{cg}(r)$ has a
similar shape to $\xi _{cc}(r)$ and $\xi _{gg}(r)$ but with an amplitude
approximately equal to their geometric mean. Croft et al. (1999) estimated 
$\xi _{cg}(r)$ in real and redshift-space from the APM galaxy and cluster
surveys and found that its shape can not be described by a single power law
and that its amplitude is almost independent of cluster richness and the 
limiting magnitude of galaxies.

In this paper we compute $\xi _{cg}(r)$ using a sample of groups derived
from the final version of the 2dFGRS and the REFLEX cluster
survey, and the galaxies from the 2MASS (Skrutskie et al. 1997) and APM
(Maddox et al. 1990) surveys. The outline of
the paper is as follows, in section \S 2 we describe the different data sets
analysed in this work. In \S 3 we describe the method used to obtain the 3D
real space cluster-galaxy cross-correlation function from the projected
cross-correlation function. In \S 4 we test the dependence of our results 
on the magnitude limit in the galaxy catalogue, the X-ray
luminosity $L_{X}$ in the REFLEX catalogue and the virial mass $M_{vir}$ in
the 2dF Galaxy Group Catalogue. In \S 5 we show a comparison with the 
results of an N-body simulation of the $\Lambda$CDM cosmological model. 
Finally, in \S 6 we present a short discussion and our main conclusions.

\section{The Data}

\subsection{Galaxy catalogues}

\begin{figure}
{\epsfxsize=8.truecm \epsfysize=8.truecm 
\epsfbox[20 150 580 710]{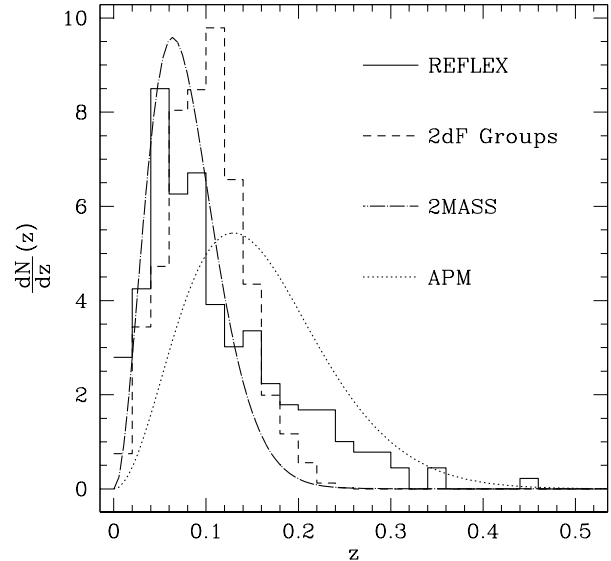}}
\caption{
Redshift distribution of the different samples of clusters and galaxies 
analysed in this work. It can be seen that the distribution of the REFLEX 
clusters peaks at approximately the same redshift as 2MASS galaxies.
}
\label{fig2}
\end{figure}

We used the information from the Two Micron All Sky Survey (2MASS; Skrutskie
et al. 1997). 2MASS characterize the large scale distribution of galaxies in
the near-infrared using the $K_{s}$ ($2.15\mu $m) passband. Our data set was
selected from the public full-sky extended source catalogue (XSC; Jarret et
al. 2000) which contains over 1.1 million extended objects brighter than 
$K_{s}=14$mag. The raw magnitudes where corrected for Galactic extinction
using the IR reddening map of Schlegel, Finkbeiner \& Davis (1998). 
There is a strong correlation between dust extinction and stellar density,
which increases exponentially towards the galactic plane. Stellar density is 
a contaminant factor of the XSC since the reliability of separating stars from
extended sources is very sensitive to this quantity (Jarret et al. 2000). In
order to avoid contamination from stars we have constructed a mask for the
2MASS survey using a $HEALPIX$ (Gorski et al. 1999) map with $Nside=256$ and
excluding those pixels where the $K_{s}$ band extinction 
$A(K_{s})=0.367\times E(B-V)>0.05$ and $\left\vert b\right\vert >20^{o}$
which reduces galactic contaminant sources to $2\%$ (Maller et al., 2003).
We also followed Maller et al. (2003) and imposed a cut at $K_{s}=13.57$ mag
in the corrected magnitudes. Figure 1 shows an Aitoff projection of the
final sample obtained with these restrictions, which contains 397447 galaxies.

We also used information from the APM Galaxy Survey (Maddox et al. 1990),
which is based on 185 UK IIIa-J Schmidt photographic plates each
corresponding to a $5.8\times 5.8$ $deg^{2}$ on the sky limited to $b_{j}\simeq
20.5$ and with a mean depth of $\simeq 400\;h^{-1}\rm{Mpc}$ for $b<-40^{o}$ and 
$\delta <-20^{o}$. We selected our galaxy sample from the APM survey imposing 
the cut $b_{j}<20.5$. The sub-sample with $b_{j}<18.5$ is also shown in 
Figure 1.

Galaxy $K_{s}$ luminosity is less sensitive to dust and stellar populations
than the $B$ band, providing a more uniform survey of the galaxy population.
The $K_{s}$ band also gives a better measure of the stellar mass content 
than the $B$ band. Then, the variations of the correlation function of 
samples selected in these bands can give us valuable information on the 
dependence of galaxy bias on the various properties of galaxies and 
then shed some light in their formation and evolution. 

\subsection{Cluster and group catalogues}

In our analysis we used information of the ROSAT-ESO X-ray Flux Limited
(REFLEX) Cluster Survey (B\"{o}hringer et al., 2004). The geometry of the 
survey is described by the southern hemisphere with $\delta < +2.5^{o}$ 
excluding the zone where $\left\vert b\right\vert <20^{o}$ and the 
Small and the Large Magellanic Clouds, covering an area of 4.24sr. 
The sample has a nominal flux limit of 
$3\times 10^{-12}\rm{erg\,s^{-1}cm^{-2}}$ 
within the ROSAT energy band (0.1-2.4)keV and comprises 447 clusters. 
Figure 1 shows an Aitoff projection of the 422 clusters inside the 2MASS 
mask described in section 2.1.

We also used a group catalogue (hereafter the 2dF Galaxy Group 
Catalogue, 2dFGGC) constructed from the final version of the 2dF Galaxy 
Redshift Survey (2dFGRS, Colless et al. 2001), using the same
technique used by Merch\'{a}n \& Zandivarez (2002) to construct the group
catalogue of the 2dFGRS 100K release (Folkes et al. 1999). This sample 
 comprises 5568 groups for which virial mass, velocity dispersion and 
virial radius have been determined.

These samples where determined using very different types of information.
The strong relation between $L_{X}$ and $M$ (see e.g. Reiprich \& 
B\"{o}hringer, 2002) indicates that X-ray selected clusters samples are 
close to be basically selected by mass. The situation is very different 
for optically selected groups and clusters where the selection criteria 
is based on richness, which is not a good mass indicator. The study of 
the differences in the way in which galaxies cluster around them can be 
very important to improve our knowledge of this type of objects and to 
understand their overall statistical properties.

Figure 2 shows the redshift distribution of the samples analysed in this
work. The distribution for 2MASS galaxies was obtained by Maller et al. 
(2004) by matching it with SDSS Early Data Release (Stoughton et al. 2002) 
and the 2dFGRS 100K release (Colles et al. 2001). For APM galaxies we 
used the fit from Baugh \& Efstathiou (1993). It can be seen that the 
distribution for the REFLEX cluster survey peaks at approximately the 
same redshift as 2MASS galaxies. As we measure the
projected cross-correlation function (see section 3), the largest
contribution to this function will come for cluster-galaxy pairs laying at
similar redshifts, which enhances our signal, on the other hand the 
differences with the distribution of APM galaxies will cause a lower 
amplitude in the projected correlations.

\section{Cluster-Galaxy Correlations}

\begin{figure*}
\centering
\centerline{\epsfysize = 12cm \epsfbox[30 160 580 670]{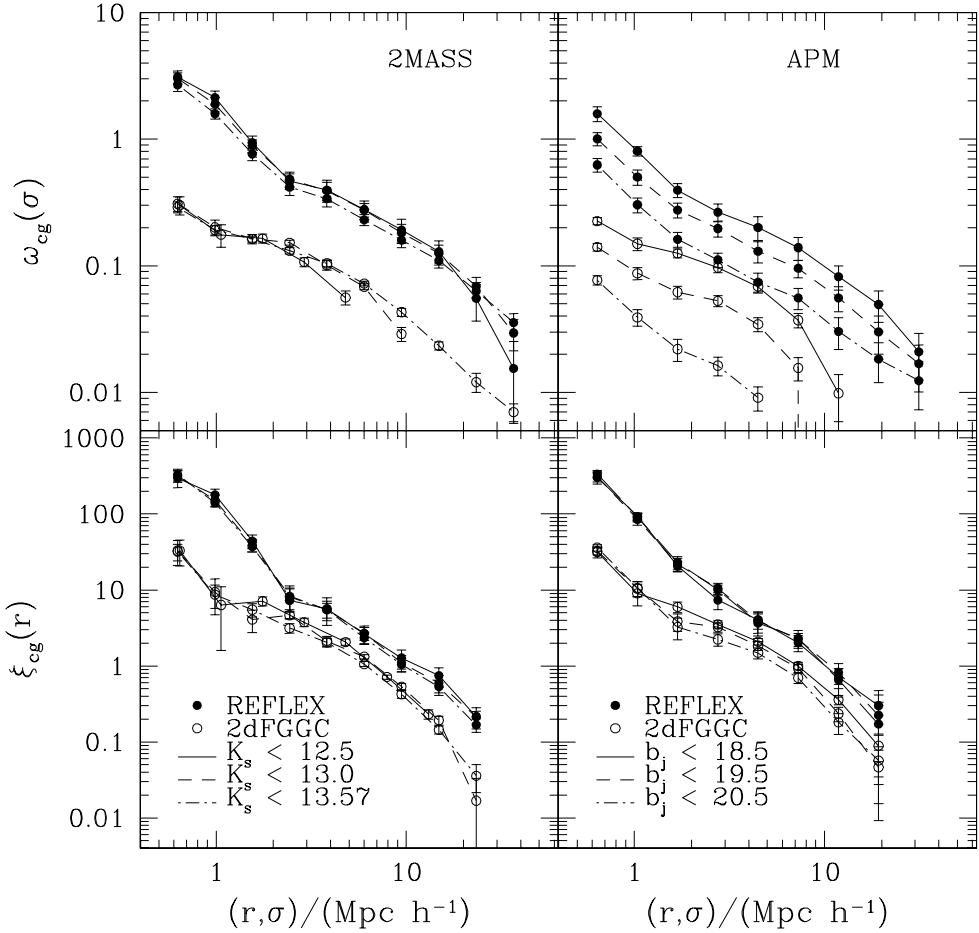}}
\caption{
The upper panels show the projected cross-correlation function 
$\omega _{cg}(\sigma )$ obtained for REFLEX clusters and 2dF groups against
APM (right) and 2MASS (left) galaxies for three different magnitude limits.
In both cases the error bars where obtained by the bootstrap re-sampling
technique. The lower panels show the correspondent 3D real-space
cross-correlation function recovered from $\omega _{cg}(\sigma )$.
}
\label{fig3}
\end{figure*}

The spatial cluster-galaxy cross-correlation function $\xi _{cg}(r)$ is
defined so that the probability $dP$ of finding a galaxy in the volume
element $dV$ at a distance $r$ from the center of a cluster is 
\begin{equation}
dP=\bar{n}\left[ 1+\xi _{cg}(r)\right] dV,  \label{definicion}
\end{equation}%
where $\bar{n}$ is the mean space density of galaxies. We only have
information on the distance (redshifts) of the clusters and not on the
individual galaxies in the 2MASS and APM catalogues. Then, in order to
obtain $\xi _{cg}(r)$ we first determine the projected cross-correlation
function $\omega _{cg}(\sigma )$, defined by Lilje \& Efstathiou (1988),
where $\sigma =cz\theta /H_{0}$ is the projected separation between a
cluster at redshift $z$ and a galaxy at angular distance $\theta $ from its
centre. To determine this quantity from our data we used the estimator 
\begin{equation}
\omega (\sigma )=\frac{\left\langle CG(\sigma )\right\rangle }{\left\langle
CR(\sigma )\right\rangle} \frac{N_R}{N_G}-1,  \label{estimador}
\end{equation}%
where $N_G$ is the number of galaxies, $\left\langle CG(\sigma )\right\rangle $ is number of cluster-galaxy pairs at a projected distance $\sigma$, and 
$\left\langle CR(\sigma )\right\rangle $ is the analogous quantity defined 
for a random distribution of $N_R$ points covering the same angular mask 
than the original galaxy survey.

When the angle $\theta \ll 1$ rad, that is, the distance $y$ to the 
cluster centre is much larger than the projected cluster-galaxy separation, 
the projected correlation function is related to the spatial correlation
function by the simple integral equation (Saunders, Rowan-Robinson 
\& Lawrence, 1992) 
\begin{equation}
\xi(r )=-\frac{1}{B\pi}\int_{r}^{\infty}\frac{d\omega_{cg}(\sigma)}{d\sigma} 
\frac{1}{(\sigma^{2}-r^{2})^{1/2}}d\sigma.  \label{integral}
\end{equation}
The constant $B$ regulates the amplitude of the correlation function taking
into account the differences in the selection function of clusters and
galaxies and can be calculated by (Lylje \& Efstathiou 1988) 
\begin{equation}
B=\frac{\sum_{i}\psi(y_i)} {\sum_{i}\frac{1}{y_{i}^2}\int_{0}^{\infty}%
\psi(x)x^2dx},  \label{factorb}
\end{equation}
where $\psi(y)$ is the selection function of the galaxy survey, $y_i$ is the
distance to cluster $i$ and the sums extends to all clusters in the sample.

Equation \ref{integral} can be easily solved analytically if we perform a
linear interpolation of $\omega _{cg}$ between its values at the measured $%
\sigma $'s. With this approach the solution to equation \ref{integral} is
(Saunders, Rowan-Robinson \& Lawrence, 1992) 
\begin{equation}
\xi _{cg}(\sigma _{i})=-\frac{1}{B\pi }\sum_{i\leq j}\frac{\omega
_{j+1}-\omega _{j}}{\sigma _{j+1}-\sigma _{j}}\ln \left( \frac{\sigma _{j+1}+
\sqrt{\sigma _{j+1}^{2}-\sigma _{i}^{2}}}{\sigma _{j}+\sqrt{\sigma
_{j}^{2}-\sigma _{i}^{2}}}\right) .  \label{solucion}
\end{equation}
We used this expression to obtain $\xi _{cg}$ from $\omega _{cg}$. In order
to calculate the factor $B$ we evaluated the selection function $\psi (y)$ by
\begin{equation}
\psi(y)=\int_{L_{min}(y)}^{\infty}\phi(L)\,\rm{dL},  \label{selecfunc}
\end{equation}
where $L_{min}(y)$ is the minimum luminosity that a galaxy at distance $y$ must 
have in order to be included in the catalogue and $\phi(x)$ is the luminosity 
function, for which we used the parameters of Bell et al. (2003) for 2MASS and 
Loveday et al. (1992) for APM.

\section{RESULTS}

\subsection{The projected and 3D real-space cross-correlation function}

The upper panels of Figure 3 show the projected cross-correlation function $%
\omega _{cg}(\sigma )$ obtained for REFLEX clusters and 2dF groups against
APM (right) and 2MASS (left) galaxies for three different magnitude limits.
In both cases the error bars where obtained by the bootstrap re-sampling
technique. Specially for 2dF groups, the shape of this function is different
from a simple power law as it shows an increment in small scales that
can be associated to the inner profile of the groups and clusters.

\begin{table*}
\begin{center}
\caption{Results for power-law fits of the form $\xi_{cg}(r)=(r/r_0)^{\gamma}$
to the inner and outer regions of  $\xi_{cg}(r)$ for the different samples
analysed using the REFLEX clusters as centres.}
\begin{tabular}{ccccccc}
\hline\hline
REFLEX sample & Galaxy sample & $r^{inner}_{0}$ & $\gamma^{inner}$& $r^{outer}_{0}$ & $\gamma^{outer}$ & \\ 
\hline\hline
all  & 2MASS $K_s <12.5$ & $6.1^{+1.1}_{-0.6}$  & $-2.60^{+0.26}_{-0.17}$ & $10.3^{+0.90}_{-1.17}$ 
                               & $-1.55^{+0.17}_{-0.19}$ & \\
all  & 2MASS $K_s <13.0$ & $6.2^{+1.1}_{-0.7}$  & $-2.52^{+0.22}_{-0.19}$ & $9.85^{+0.67}_{-0.73}$  
                               & $-1.69^{+0.09}_{-0.14}$ & \\
all  & 2MASS $K_s <13.57$ & $5.97^{+1.30}_{-0.46}$  & $-2.62^{+0.19}_{-0.15}$   & $9.18^{+0.60}_{-0.59}$ 
                               & $-1.77^{+0.13}_{-0.12}$ & \\
\hline
all  & APM $b_j <18.5$ & $5.47^{+0.64}_{-0.38}$  & $-2.69^{+0.24}_{-0.10}$   & $9.9^{+1.0}_{-1.1}$  & 
                          $-1.67^{+0.16}_{-0.15}$ & \\
all  & APM $b_j <19.5$ & $5.47^{+1.0}_{-0.4}$  & $-2.61^{+0.25}_{-0.17}$   & $10.4^{+1.3}_{-1.1}$  & 
                          $-1.71^{+0.13}_{-0.14}$ & \\
all  & APM $b_j <20.5$ & $5.54^{+0.88}_{-0.38}$  & $-2.65^{+0.21}_{-0.14}$ & $9.5^{+1.2}_{-1.0}$  & 
                          $-1.79^{+0.16}_{-0.19}$ & \\
\hline
$\log(L_{X}/10^{44}\rm{erg\ s^{-1}})>-1.5$ & 2MASS $K_s <13.57$ & $5.96^{+0.79}_{-0.46}$ & 
                  $-2.62^{+0.19}_{-0.15}$ & $9.18^{+0.59}_{-0.59}$ & $-1.77^{+0.13}_{0.13}$  &   \\
$\log(L_{X}/10^{44}\rm{erg\ s^{-1}})>-1.0$ & 2MASS $K_s <13.57$ & $6.24^{+0.95}_{-0.61}$ & 
                  $-2.57^{+0.18}_{-0.18}$ & $10.28^{+0.67}_{-0.66}$ & $-1.70^{+0.14}_{-0.13}$ &   \\
$\log(L_{X}/10^{44}\rm{erg\ s^{-1}})>-0.85$ & 2MASS $K_s <13.57$ & $6.24^{+0.83}_{-0.57}$ & 
                  $-2.58^{+0.21}_{-0.22}$ & $10.45^{+0.58}_{-0.65}$ & $-1.75^{+0.10}_{-0.12}$ &   \\
$\log(L_{X}/10^{44}\rm{erg\ s^{-1}})>-0.5$ & 2MASS $K_s <13.57$ & $6.41^{+0.45}_{-0.27}$ & 
                  $-2.75^{+0.11}_{-0.06}$ & $10.8^{+1.2}_{-0.7}$ & $-1.75^{+0.11}_{-0.12}$ &   \\
$\log(L_{X}/10^{44}\rm{erg\ s^{-1}})>0$ & 2MASS $K_s <13.57$    & $7.13^{+0.89}_{-0.44}$ & 
                  $-2.72^{+0.27}_{-0.07}$ & $13.1^{+1.1}_{-1.1}$ & $-1.69^{+0.15}_{-0.18}$ &   \\

\hline\hline
\end{tabular}
\end{center}

\end{table*}

\begin{table*}
\begin{center}
\caption{Results for power-law fits of the form $\xi_{cg}(r)=(r/r_0)^{\gamma}$
to the inner and outer regions of  $\xi_{cg}(r)$ for the different samples
analysed using 2dF groups as centres.}
\begin{tabular}{ccccccc}
\hline\hline
2dF Groups sample& Galaxy sample & $r^{inner}_{0}$ & $\gamma^{inner}$& $r^{outer}_{0}$ & $\gamma^{outer}$ & \\ 
\hline\hline
all  & 2MASS $K_s <12.5$ & $1.9^{+1.6}_{-0.5}$  & $-3.28^{+0.32}_{-0.26}$ & $6.28^{+0.20}_{-0.22}$ 
                               & $-1.60^{+0.07}_{-0.07}$ & \\
all  & 2MASS $K_s <13.0$ & $2.2^{+1.5}_{-0.6}$  & $-2.78^{+0.34}_{-0.35}$ & $6.21^{+0.21}_{-0.21}$  
                               & $-1.51^{+0.06}_{-0.06}$ & \\
all  & 2MASS $K_s <13.57$ & $2.1^{+1.2}_{-0.3}$  & $-2.90^{+0.25}_{-0.40}$ & $5.18^{+0.18}_{-0.20}$ 
                               & $-1.58^{+0.06}_{-0.06}$ & \\
\hline
all  & APM $b_j <18.5$ & $1.8^{+1.1}_{-0.2}$  & $-1.75^{+0.50}_{-0.61}$   & $6.65^{+0.33}_{-0.37}$  & 
                          $-1.37^{+0.08}_{-0.08}$ & \\
all  & APM $b_j <19.5$ & $2.6^{+1.3}_{-0.7}$  & $-2.02^{+0.59}_{-0.69}$   & $5.71^{+0.35}_{-0.35}$  & 
                          $-1.42^{+0.09}_{-0.09}$ & \\
all  & APM $b_j <20.5$ & $2.5^{+1.1}_{-0.6}$  & $-2.20^{+0.43}_{-0.61}$ & $4.82^{+0.34}_{-0.36}$  & 
                          $-1.46^{+0.12}_{-0.11}$ & \\
\hline
 $\log(M_{vir}/M_{\odot })>13$ & 2MASS $K_s <13.57$ & $1.95^{+0.53}_{-0.16}$ & 
                  $-3.23^{+0.82}_{-0.55}$ & $6.73^{+0.27}_{-0.25}$ & $-1.52^{+0.06}_{0.07}$  &   \\
 $\log(M_{vir}/M_{\odot })>13.5$ & 2MASS $K_s <13.57$ & $2.29^{+0.65}_{-0.13}$ & 
                  $-3.47^{+0.94}_{-0.31}$ & $8.21^{+0.39}_{-0.35}$ & $-1.59^{+0.08}_{-0.07}$ &   \\
 $\log(M_{vir}/M_{\odot })>14$ & 2MASS $K_s <13.57$ & $5.7^{+1.2}_{-1.1}$ & 
                  $-2.47^{+0.29}_{-0.28}$ & $10.6^{+1.0}_{-1.1}$ & $-1.63^{+0.15}_{-0.17}$ &   \\
 $\log(M_{vir}/M_{\odot })>14.5$ & 2MASS $K_s <13.57$   & $6.3^{+1.5}_{-1.5}$ & 
                  $-2.74^{+0.47}_{-0.42}$ & $10.8^{+1.7}_{-1.5}$ & $-1.84^{+0.35}_{-0.35}$ &   \\

\hline\hline
\end{tabular}
\end{center}

\end{table*}

The lower panels of Figure 3 show the corresponding 3D real-space
cross-correlation function recovered from $\omega _{cg}(\sigma )$. 
Our findings are in complete agreement with the results of Croft et al. (1999, 
see their figure 2). We found that the shape of the correlation function can 
not be described as a single power law for all scales, as it becomes steeper 
at small scales. Tables 1 and 2 show the results of power-law fits of the form
$\xi(r)_{cg}=(r/r_0)^{\gamma}$ to the inner and outer regions for the different 
samples analysed. It can be clearly seen that the correlation function is nearly
independent of the magnitude limit imposed to the galaxy sample showing 
only small variations in the best fit power-law parameters.
The shape of the cross-correlations obtained for 2MASS and APM galaxies 
are different. The feature at $r\simeq 2\;h^{-1}\rm{Mpc}$ that show the 
transition to the inner profile of the clusters is more evident for 2MASS 
galaxies. This effect can be associated with differences in the spatial 
distribution of these objects around clusters. 2MASS galaxies where selected 
in the near-infrared and so would tend to be luminous and of early type 
morphologies, and so, are expected to have a stronger clustering 
(Norberg et al. 2002). 

\subsection{Dependence on $L_{X}$ and $M_{vir}$}

In order to analyse the dependence of the distribution of galaxies around
clusters and groups on properties such as their mass or X-ray luminosity we
have computed $\xi _{cg}(r)$ for 2MASS galaxies and the REFLEX cluster
survey for four different limits on $L_{X}$. Analogously for 2dFGGC we have 
adopted four different limits in $M_{vir}$. The results obtained for these 
samples are shown in figures 4 and 5. We find a clear dependence of the amplitude 
of $\xi _{cg}(r)$ on $L_{X}$ which indicates that galaxies are more clustered
around more luminous (massive) objects. The value of the amplitude of the
correlaction function for the outer region changes from $r_0=9.18 \pm 0.59$ 
for $\log(L_{X}/10^{44}\rm{erg\ s^{-1}})>-1.5$ to $r_0=13.1 \pm 1.1$ for 
$\log(L_{X}/10^{44}\rm{erg\ s^{-1}})>0$ and the inner region shows a similar 
behaviour. The shape of $\xi _{cg}(r)$ does not show any important differences, 
but the small changes of $\gamma^{inner}$ show a weak indication that a higher 
limiting $L_{X}$ produce steeper inner regions.

The changes in $\xi _{cg}(r)$ for 2dF groups and 2MASS varying the
limiting virial mass are even stronger. The overall shape of $\xi _{cg}(r)$
changes, specially for small scales where the amplitude increase from the
sample with $\log(M_{vir}/M_{\odot })>13$ to the one with 
$\log(M_{vir}/M_{\odot })>14.5$ is a factor 20 whereas for scales $r\simeq
10\;h^{-1}\rm{Mpc}$ the increase is a factor 5. The value of $\gamma$ in the 
outer region also increase with $M_{vir}$ and for $\log(M_{vir}/M_{\odot })>14.5$
its value is simmilar to the one in the inner region.  

\begin{figure}
{\epsfxsize=8.truecm \epsfysize=8.truecm 
\epsfbox[20 150 580 710]{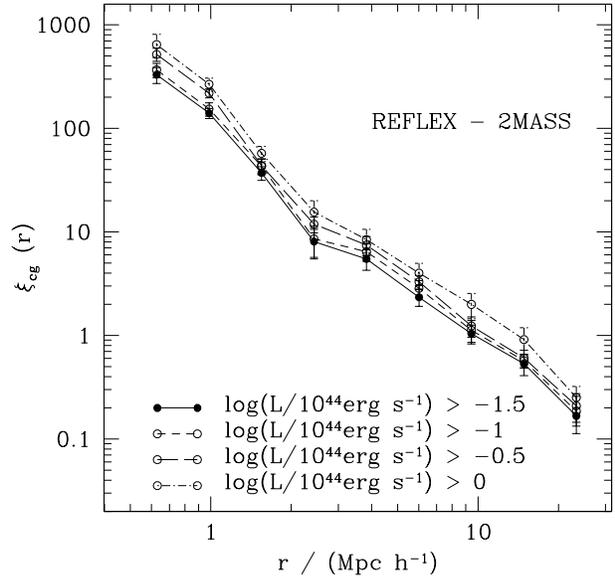}}
\caption{
The real space cluster-galaxy cross-correlation functions of REFLEX clusters 
and all 2MASS galaxies for different X-ray luminosity limits. 
}
\label{fig4}
\end{figure}

\begin{figure}
{\epsfxsize=8.truecm \epsfysize=8.truecm 
\epsfbox[20 150 580 710]{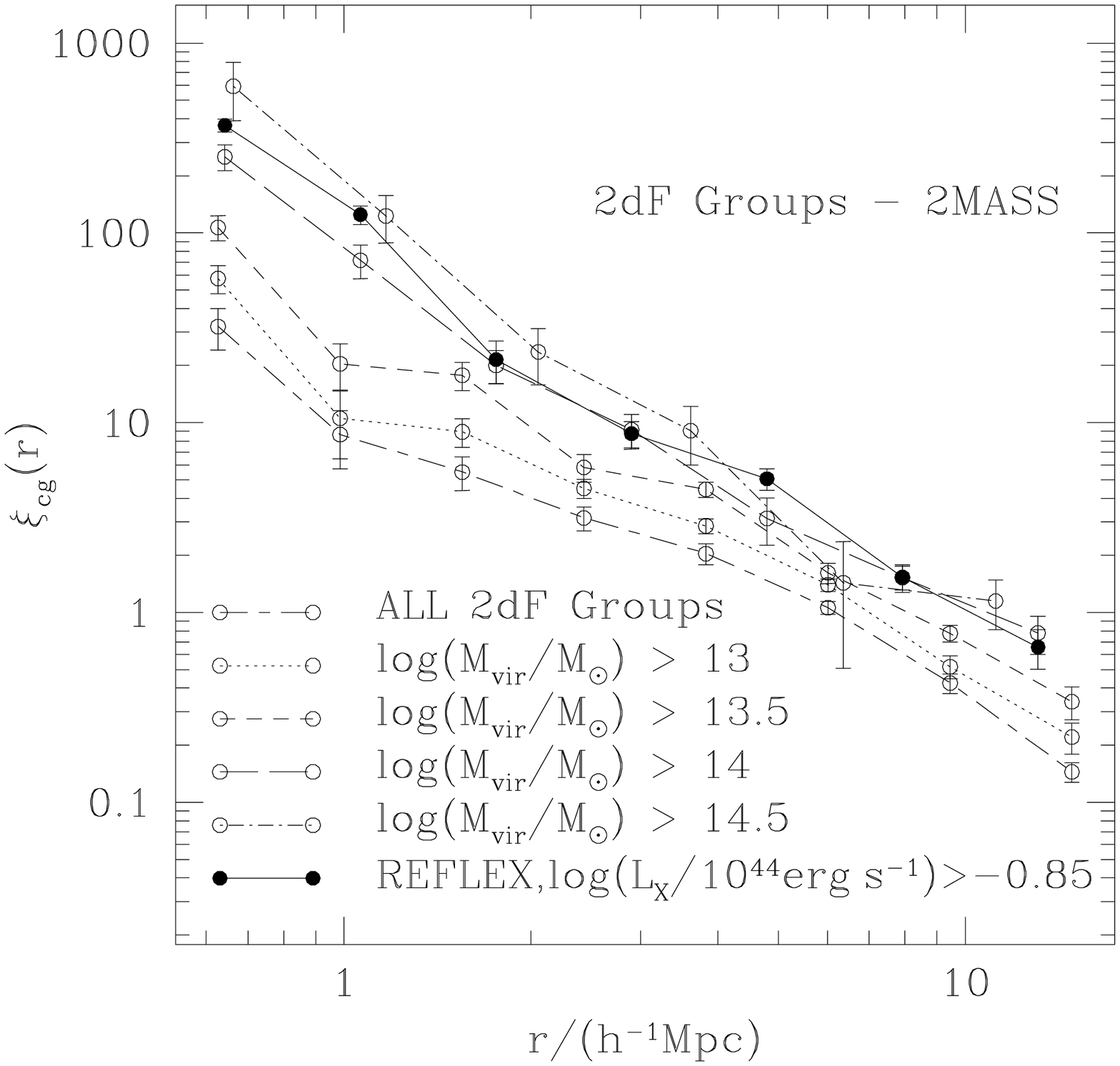}}
\caption{
The real space cluster-galaxy cross-correlation functions of 2dFGGC 
and all 2MASS galaxies for different limits on $M_{vir}$. For comparison 
$\xi_{cg}(r)$ for REFLEX clusters with 
$\log(L_{X}/10^{44}\rm{erg\ s^{-1}}) > -0.85$ and 2MASS galaxies 
is also shown.
}
\label{fig5}
\end{figure}

Figure 5 also shows for comparison $\xi _{cg}(r)$ for REFLEX clusters with 
$\log(L_{X}/10^{44}\rm{erg\ s^{-1}})>-0.85$. Its similarity with the 
cross-correlation function obtained using 2dF groups with 
$\log(M_{vir}/M_{\odot })>14$ is remarkable.
This fact can be used as an indirect test of the $L_{X}-M$ relation.
Reiprich \& B\"{o}hringer (2002) found a power law relation between $L_{X}$ 
in the {\it ROSAT} energy band (0.1-2.4 keV) and the mass $M_{200}$ 
(where $M_{200}=M_{total}(r<r_{200})$) of the form 
\begin{equation}
\log \left[ \frac{L_{X}}{h_{50}^{-2}10^{40}\rm{erg\ s^{-1}}}\right] =A+\alpha
\log \left( \frac{M_{200}}{h_{50}^{-1}M_{\odot }}\right) .  \label{lx-m}
\end{equation}%
The transformation from $M_{vir}$ to $M_{200}$ can be done assuming a given
density profile for the clusters (see White 2000). For a $\Lambda$CDM the 
difference is small ($M_{vir} \simeq 1.2 M_{200}$). Then, assuming 
 $h=0.7$, $A=-20.055$ and $\alpha = 1.652$ (from table 7 in Reiprich and 
B\"ohringer, 2002), $\log(M_{vir}/h^{-1}M_{\odot })=14.0$ corresponds to 
$\log (L_{X}/10^{44}\rm{erg\ s^{-1}})=-0.85$ in excellent agreement with the 
limit imposed to the sample used in figure 5.

\section{Comparison with the results of $\Lambda$CDM N-body simulations}

\begin{figure*}
\centering
\centerline{\epsfysize = 12cm \epsfbox[30 160 580 700]{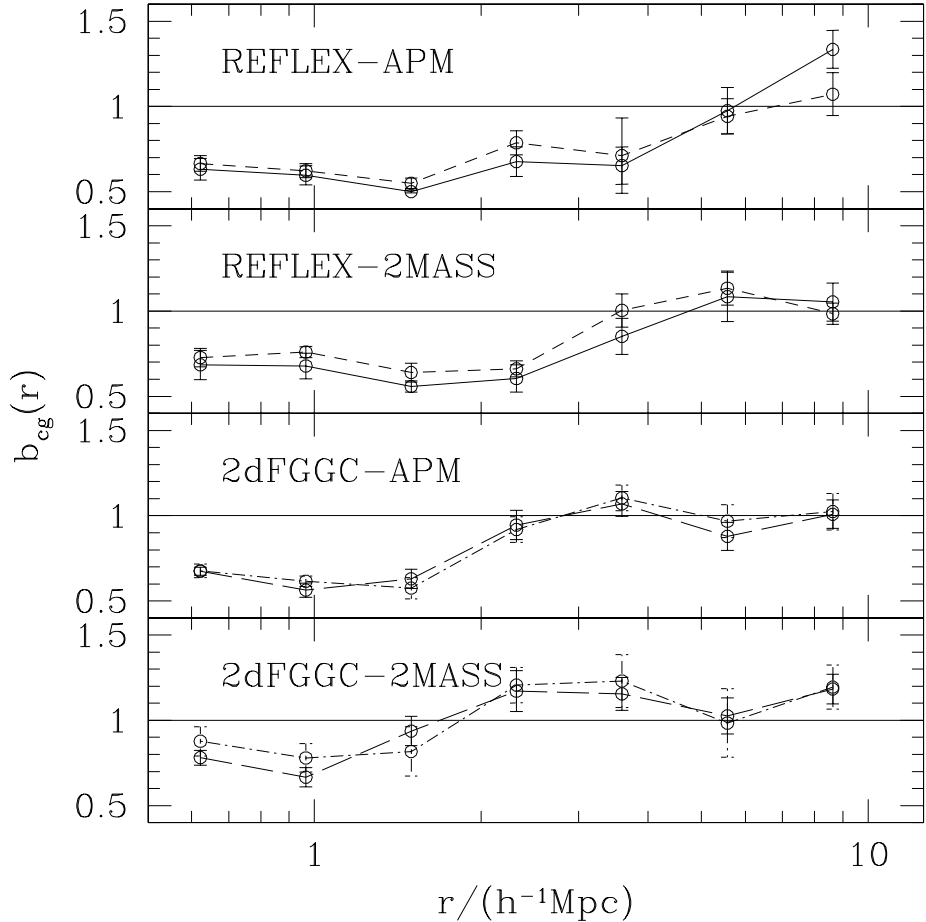}}
\caption{
The results for $b_{cg}(r)$ for all the samples analysed in this work 
obtained by the comparison with the results of a $\Lambda$CDM N-body 
simulation. Different lines show the results obtained for samples selected 
with different limiting mass given by $m_1 = 7.5 \times 10^{13}\,M_{\odot}$ 
(dot-dashed), $m_2 = 1.4 \times 10^{14}\,M_{\odot}$ (long dashed), 
$m_3 = 4.1 \times 10^{14}\,M_{\odot}$ (short dashed) and $m_4 = 7.5 
\times 10^{14}\,M_{\odot}$ (solid). 
}
\label{fig6}
\end{figure*}

In order to make a comparison between our results and those corresponding to
the concordance cosmological model we used the $\Lambda $CDM Very Large
Simulations (VLS). This simulations where carried out by the Virgo
Supercomputing Consortium using computers based at Computing Centre of the
Max-Planck Society in Garching and at the Edinburgh Parallel Computing
Centre. The data are publicly available at www.mpa-garching.mpg.de/NumCos.
It uses $512^{3}$ particles in a box of $480\;h^{-1}\rm{Mpc}$, with initial
conditions consistent with a $\Lambda$CDM power spectrum computed using 
CMBFAST (Seljak \& Zaldarriaga, 1996), normalized so that $\sigma _{8}=0.9$. 
We have identified groups using a Friends-Of-Friends (FOF) algorithm with 
a linking parameter $b=0.2$ and we used these halos as centres to measure the
halo-mass cross-correlation function $\xi _{hm}(r)$ for different values of
limiting mass. We used these results to infer the scale dependence of the
cluster-galaxy bias factor, that is the bias between the distribution of
galaxies and mass around clusters, defined by 
\begin{equation}
b_{cg}(r)=\xi _{cg}(r)/\xi _{hm}(r).  \label{bias}
\end{equation}%
For small scales this is simply the difference between the density profile
of mass and galaxies. 

  In order to obtain $b_{cg}(r)$ we used four values of the limiting mass 
to measure $\xi _{hm}(r)$: $m_1 = 7.5 \times 10^{13}\,M_{\odot}$, $m_2 = 1.4 
\times 10^{14}\,M_{\odot}$, $m_3 = 4.1 \times 10^{14}\,M_{\odot}$ and 
$m_4 = 7.5 \times 10^{14}\,M_{\odot}$. As the REFLEX clusters and the 2dF 
groups span different ranges of mass, we used the 2dFGGC as centres to 
calculate $\xi _{cg}(r)$ for $m_1$ and $m_2$, and the REFLEX catalogue for 
$m_3$ and $m_4$. When using the 2dFGGC we simply imposed the same limits 
in $m_{vir}$. For the REFLEX clusters, we used the X-ray luminosity 
threshold for which the abundance, calculated by the integration of the 
X-ray luminosity function of B\"{o}hringer et al. (2002), gives the same 
values as those obtained in the simulation. The results obtained are shown 
in Figure 6. 
  
   Our results for the two lower values of the threshold mass show that, 
at large scales, $b_{cg}(r)$ is consistent with a constant of order 
$b_{cg}\simeq 1$ for APM galaxies and slightly higher for 2MASS, which 
shows that 2MASS galaxies are more biased tracers of the mass distribution 
than optically selected galaxies. On smaller scales ($r < 2\;h^{-1}\rm{Mpc}$) 
the distribution of galaxies around clusters shows a significant and 
nearly constant anti-bias of $b_{cg}\simeq 0.7$. For $m_3$ and $m_4$, 
$b_{cg}(r)$ shows the same behavior but with a larger 
transition scale between the two regimes ($r \simeq 2\;h^{-1}\rm{Mpc}$). In all 
cases this scale would correspond to the onset of non-linearities, where 
$\xi _{cg}(r)$ becomes steeper and virialization takes place. A constant 
bias is a feature characteristic of linear theory and then it is not 
surprising that it does not hold as well in the non-linear 
regime. Another reason that may add to the anti-bias in the cluster and 
infall regions is the change in the galaxy types. The dense environments 
contain galaxies with older stellar populations with less light per 
baryonic mass. Then, cluster regions are down-weighted by this effect.  
It is very important to note that these results are independent of the 
limiting values of mass, X-ray luminosity and magnitude used to construct 
the clusters and galaxy samples, which shows that this is not a property of 
a given galaxy type, but a generic feature of the processes that control the 
efficiency of galaxy formation and evolution. 

\section{Discussion and Conclusions}

In this work we have performed the first detailed calculation of the 3D
real-space cluster-galaxy cross-correlation for an X-ray selected cluster
sample and galaxies selected in the optical and near-infrared wave-bands. We
have also presented the first calculation of this statistical measure for a
galaxy group sample from the 2dFGRS. These two samples span a wide range
of masses and so they allow us to analyse the dependence of $\xi _{cg}(r)$ on 
$M_{vir}$. Our results for the different sub-samples of galaxies from 2MASS 
or APM with different limiting magnitudes show that $\xi _{cg}(r)$ is 
almost independent of galaxy properties and that its shape is determined 
almost exclusively by the criteria used to define the cluster sample.

In agreement with previous works (Croft et al. 1999) we found that the shape
of $\xi _{cg}(r)$ can not be described by a single power law at all scales.
Instead, it shows two regimes with a clear transition, one for the larger
scales, and a steeper one at smaller scales showing the inner profiles of
the distribution of galaxies around cluster centres. We have used our 
results to check the $L_{X}-M$ relation of Reiprich \& B\"{o}hringer (2002). 
We have found that the correlation functions obtained using 2dF groups with 
a given lower mass limit and REFLEX clusters with the corresponding limit 
on $L_{X}$ are in complete agreement showing the validity of this relation.

The comparison of our results with those obtained for the 
halo-mass cross-correlation function in a $\Lambda $CDM N-body
simulation shows that the observational results are consistent with a
constant bias on large scales of order unity for APM galaxies and slightly 
higher for 2MASS galaxies. On smaller scales our results suggests that 
there is a substantial anti-bias ($b_{cg}(r)\simeq0.7$). In all cases the 
transition scale between the two regimes corresponds to the onset of 
non-linearities, where the correlation functions becomes steeper. This is 
a strong result of our analysis which is independent of the properties 
of the cluster or galaxy samples used, showing that it is simply a 
generic feature of the different processes that govern galaxy formation 
and their subsequent evolution.

\section*{Acknowledgments}

We thank Maximiliano Pivato for helping us with the analysis of the VLS
simulations. AGSV thanks the hospitality of the Max-Planck-Institut f\"{u}r
Extraterrestriche Physik during his visit. This publication makes use of
data products from the Two Micron All Sky Survey, which is a joint project
of the University of Massachusetts and the Infrared Processing and Analysis
Center/California Institute of Technology, funded by the National
Aeronautics and Space Administration and the National Science Foundation.
This work was partially supported by the Concejo Nacional de Investigaciones
Cient\'{\i}ficas y Tecnol\'{o}gicas (CONICET), the Secretar\'{\i}a de
Ciencia y T\'{e}cnica (UNC) and the Agencia C\'{o}rdoba Ciencia.

\end{document}